\documentclass[preprint,12pt]{elsarticle}
\usepackage{subeqn}

\usepackage{amssymb}

 \biboptions{sort&compress}
\journal{Optics Communications}

\begin{document}

\begin{frontmatter}



\title{Nonparaxial spatial optical solitons with nonlocality nonlinearity}

\author{Jilong Chen}
\author{Zhiwei Shi}
\address{$^1$Faculty of Information Engineering, Guangdong
University of Technology, Guangzhou 510006,P.R.China}

\author{Huagang Li}
\address{$^2$Department of Physics, Guangdong University of
Education, Guangzhou 510303, P.R.China}
\ead{lihuagang@gdei.edu.cn}

\begin{abstract}
We investigate one-dimensional nonparaxial spatial optical solitons with nonlocal nonlinearity. We show an exact analytical solution to the nonlocal nonlinear nonparaxial propagation equation in the cases of high and weak nonlocality.
We also numerically find that the degree of nonlocality can affect the width of nonlocal soliton beams, but have no effect on their stability. Contrarily, nonparaxiality can affect their stability, but have no effect on their width.
\end{abstract}


\end{frontmatter}



\section{Introduction}
A nonlocal nonlinear response played an important role on the optical spatial solitons over the years. Nonlocality exists in different physical media, as nematic liquid crystals~\cite{1}, photorefractive media~\cite{2}, thermal~\cite{3,4}, and so on. Various degree of nonlocality given by the width of the nonlocal response function $R(x)$ and the intensity profile of the beam can be divided into four types, like local, weakly nonlocal, general nonlocal and highly nonlocal response~\cite{9}.

Nonparaxiality can refer to two different
contexts of distinct nonparaxial character: high intensity and large angles of propagation~\cite{10}.
The first type of nonparaxiality results from the evolution of ultra-narrow beams in nonlinear media. In 1993, Akhmediev et al cast doubt on the suitability and limitations of the normalized nonlinear Schr\"odinger (NLS) equation for describing the evolution of such beams~\cite{10-1}. The scalar theory of the self-focusing of an optical beam is not valid for a very narrow beam, and a vector nonparaxial theory is
developed from the vector Maxwell equations~\cite{11,12}. Based on these equations, Crosignani et al have reported and analysed bright~\cite{13} and
dark~\cite{14} nonparaxial solitons. In contrast to the first type, the second type of nonparaxiality arises from the rapid evolution of the field envelope of a broad (when compared
to the wavelength) beam propagating at a large angle to the longitudinal axis~\cite{10}. The
scalar nonlinear Helmholtz (NLH) equation well describe this nonparaxiality and overcome the limitations of the NLS~\cite{16}. Exact analytical soliton
solutions have been found in a focusing Kerr type medium~\cite{19}. Nonparaxial theory based on NLH equation has also been applied to find dark Kerr~\cite{20}, two-component~\cite{21}, boundary~\cite{22}, and
bistable~\cite{23} Helmholtz soliton solutions. At nonlinear interfaces, Soliton refraction effects have a strong inherent
angular character and constitute an excellent
testbed for nonparaxial Helmholtz theory~\cite{10,25,26,27}.

 In this paper, we investigate the nonparaxial solitons in the nonlocal nonlinear medium. We find an exact analytical solution to the nonlocal nonlinear nonparaxial propagation equation in the cases of high and weak nonlocality. In addition, we numerically study the influence of the degree of nonlocality and nonparaxiality on the width and stability of solitons. We devote Section \ref{Theory model} to a detail description of the theory model of the nonlocal Helmholtz soltions. Section \ref{Exact analytical soltions solutions} studies the exact analytical nonlocal Helmholtz solitons solutions, which correspond to the high and weak nonlocality. Section
\ref{Numerical solitons solutions} is devoted to studying the numerical solitons soluitions. The aim is to illustrate the influence of the degree of nonlocality, nonparaxiality on the width and stability of solitons. Section \ref{Conclusion} summarizes the main conclusions of the paper.

\section{Theoretical model}

\label{Theory model}
For the simple one-dimensional case, the optical field satisfies the Helmholtz equation
\begin{equation}
\frac{\partial^2 \vec{E}}{\partial Z^2}+\frac{\partial^2 \vec{E}}{\partial X^2}+k_{0}^{2}n^2(\vec{E})\vec{E}=0,
\label{eq:one}
\end{equation}
where $k_{0}=\omega/c$ is the propagation constant in vacuum, $c$ is the speed of light, and $\omega$ is the frequency.
Introducing a normalization appropriate to a forward propagating beam, $\vec{E}(X,Z)=A(X,Z)exp(ikZ)$, a nonlocal nonlinearity $n(\vec{E})=n_0+\Delta n$ and
assuming that the approximation $n^2(\vec{E})=n_0^2+2n_0\Delta n$, where $k$ is the wave number in the medium, and $\Delta n$ is the nonlinear induced change of the refractive index, which satisfies
\begin{equation}
w_m^2\frac{\partial^2 \Delta n}{\partial X^2}-\Delta n+n_2|\vec{E}|=0,
\label{eq:two}
\end{equation}
where $n_2$ is the nonlinear coefficient. We have employed the following normalizations, $z=Z/L_D$, $x=\sqrt{2}X/w_0$, $u(x,z)=\sqrt{kn_2L_d/n_0}A(x,z)$, $\Delta n=n_0\phi/(kL_D)$, $d_0=d/\sqrt{2}$.
$w_0$ is a transverse scale parameter that we shall later relate to the width of nonparaxial
soliton beams. This scale parameter can also be considered as equivalent
to the waist of a (reference) paraxial Gaussian beam, at $z=0$, which has a
diffraction length $L_D=kw^2_0/2$. $w_m$ is the characteristic length of the nonlinear response, and $d=w_m/w_0$ stands for the degree of nonlocality of the nonlinear response. So, we get the following non-paraxial nonlinear Schr\"{o}dinger equation(NNSE) for the dimensionless amplitude $u$ of the light field coupled to the equation for normalized nonlinear induced change of the refractive index $\phi$
\begin{subequations}
\label{eq:three}
\begin{equation}
\kappa\frac{\partial^2 u}{\partial z^2 }+i\frac{\partial u}{\partial z}+\frac{1}{2}\frac{\partial^2 u}{\partial x^2 }+\phi u=0,\label{3a}
\end{equation}
\begin{equation}
d_0^{2}\frac{\partial^{2}\phi}{\partial x^{2}}-\phi+|u|^{2}=0,\label{3b}
\end{equation}
\end{subequations}
where $\kappa=1/(k^2w^2_0)$ is the non-paraxial paremeter of the NNSE. In the limits $\kappa\rightarrow0$ and $d\rightarrow0$, the nonlinear  Schr\"{o}dinger equation(NSE) can be recovered from the system (\ref{eq:three}) which describes a local nonlinear response at $d\rightarrow0$ and a strongly nonlocal response at $d\rightarrow\infty$. For Eq.(~\ref{3b}), we can also write it into the form of convolution,
\begin{equation}
\label{eq:four}
\phi=\int^{+\infty}_{-\infty}R(x-x^{'})I(x^{'})dx^{'},
\end{equation}
where $I=I(x,z)=|u(x,z)|^2$, $R(x)=1/(\sqrt{2}d)\exp(-\sqrt{2}|x|/d)$.

\section{Exact analytical soltions solutions}
\label{Exact analytical soltions solutions}
\begin{figure}[htb]
\centerline{\includegraphics[width=10cm]{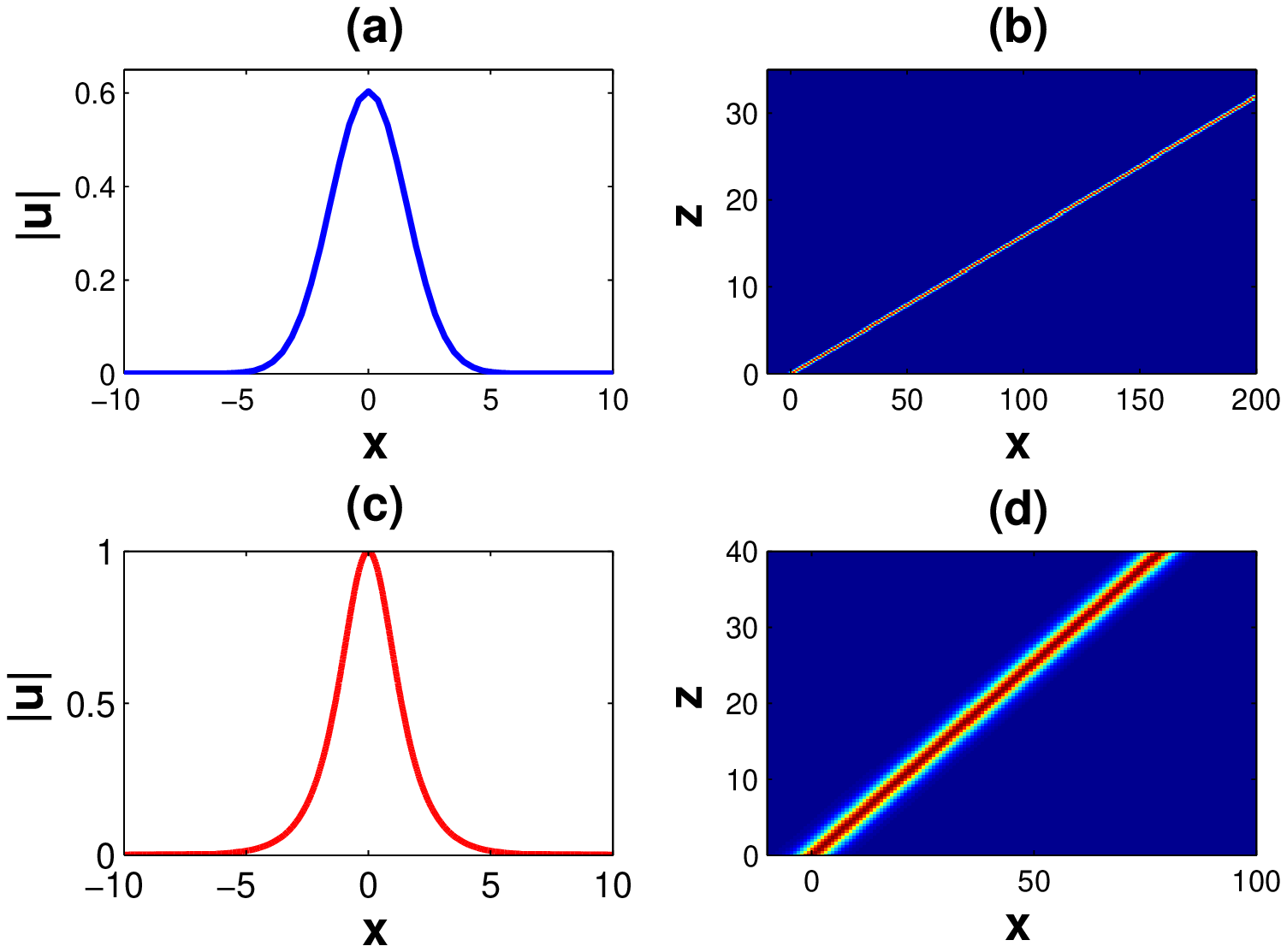}}
\caption{\label{fig:one}(Color online) (a) The profile of the exact soliton solution when $d=2.01$.
(b) Simulated propagation of the beam at $d=2.01$ with $\rho=2\%$ noise for the exact soliton solution initial condition.
(c) The profile of the exact soliton solution when $d=0.11$. (b) Simulated propagation of the beam at $d=0.11$ with $\rho=2\%$ noise for the exact soliton solution initial condition. The other parameters $\kappa=1\times10^{-3}$ and $V=10$.}
\end{figure}

When the nonlocal response is high, i.e., when the response function
$R(x)$ is wider compared with the extent of the beam. So, to find exact
analytical solutions for the spatial solitons, we expand $R(x^{'},z)$ around the point $x^{'}=x$ and combine Eqs.~(\ref{3a}) and ~(\ref{eq:four}) to obtain~\cite{27-0}
\begin{equation}
\kappa\frac{\partial^2 u}{\partial z^2 }+i\frac{\partial u}{\partial z}+\frac{1}{2}\frac{\partial^2 u}{\partial x^2 }+\frac{P_0}{\sqrt{2}d}u-\frac{1}{2}\mu P_0 x^2u=0,\label{eq:five1}
\end{equation}
where $\mu=-R_0^{''}>0$ [$R_0^{''}=\partial^2_xR(x)|_{x=0}$, $R_0^{''}<0$ because $R_0$ is a maximum of $R(x)$], $P=\int|u(x,z)|^2dx$ is the beam power, and $P_0$ is the input power at $z=0$. We search for a solution to Eq.~(\ref{eq:five1}) of the Gaussian function form~\cite{27-0}
\begin{equation}
u(x,z)=\frac{\sqrt{P_0}exp[i\theta(z)]}{\sqrt{\sqrt{\pi}w(z)}}exp[-\frac{x^2}{2w(z)^2}+i\tau(z)x^2],\label{eq:five2}
\end{equation}
where $\theta$ is the phase of the complex amplitude of the solution, $w$ is beam width, $\tau$ represents the phase-front curvature
of the beam, and they are all allowed to vary with propagation distance $z$. The real amplitude of the solution has the
form $\sqrt{P_0}/\sqrt{\sqrt{\pi}w}$, owing to the conservation of the power. Inserting the trial function above into Eq.~(\ref{eq:five1}), we obtain that
$\tau=0$, $w=w_0=1/(\mu^{1/4}P_0^{1/4})$, and $\theta=(-dw_0^2\pm\sqrt{d^2w_0^4-2d^2w_0^2\kappa+2\sqrt{2}dw_0^4\kappa P_0})/(2dw_0^2\kappa)$. So, we get the exact solitons solutions. When $d=2.01$, the solution is shown in Fig.~\ref{fig:one} (a). To illustrate the stability of the soliton, we do the direct numerical simulations of Eq.~(\ref{eq:five1}) with input condition $u|_{z=0}=v(1+\rho)\exp(-iVx)$, where $\rho(x)$ is a broadband random perturbation, $v$ is the nonlocal nonparaxial soliton solution as an initial condition, and $V$ denotes transverse velocity parameter. From Fig.~\ref{fig:one} (b), we can easily see that the soliton is stable.

When the nonlocal response is weak, i.e., when the response function
$R(x)$ is narrower compared with the extent of the beam. So, to find exact
analytical solutions for the spatial solitons, we expand $I(x^{'},z)$ around the point $x^{'}=x$ and combine Eqs.~(\ref{3a}) and ~(\ref{eq:four}) to obtain
\begin{equation}
\kappa\frac{\partial^2 u}{\partial z^2 }+i\frac{\partial u}{\partial z}+\frac{1}{2}\frac{\partial^2 u}{\partial x^2 }+(|u|^2+\gamma\frac{d^2|u|^2}{dx^2})u=0,\label{eq:six}
\end{equation}
where $\gamma=1/2\int^{+\infty}_{-\infty}R(x)x^2dx$. For
weakly nonlocal meda $\gamma\ll1$ is a small nonlocality parameter. We search for a stationary soliton solution to Eq.~(\ref{eq:six}) of the form $u(x,z)=v(x)\exp(i\beta z)$,
where the profile $v(x)$ is real, symmetric, and exponentially
localized and the propagation constant $\beta$ is positive~\cite{28}.
Substituting this solution into Eq.~(\ref{eq:six}), we can get
\begin{equation}
\frac{\partial^2 v}{\partial x^2 }+2(v^2+\kappa \beta^2-\beta)v+2\gamma v\frac{\partial^2 v^2}{\partial x^2 }=0,\label{eq:seven}
\end{equation}
Eq.~(\ref{eq:seven}) can be finally integrated to give the equation~\cite{28}
\begin{equation}
\pm x=\frac{1}{v_0}\tanh^{-1}(\frac{\sigma}{v_0})+2\sqrt{\gamma}\tan^{-1}(2\sqrt{\gamma}\sigma),\label{eq:eight}
\end{equation}
where $v_0^2=2(\beta-\kappa\beta^2)$, $\sigma^2=(v_0^2-v^2)/(1+4\gamma v^2)$. This implicit relation
gives the profile of nonlocal nonparaxial spatial solitons propagating in
weakly nonlocal Kerr-like media. when $\gamma=0$ and $\kappa=0$, i.e., for the paraxial case with local nonlinearity, $v(x)=v_0$sech$(v_0x)$. The exact soliton solution is shown in Fig.~\ref{fig:one} (c) when $d=0.11$. Using Eq.~(\ref{eq:seven}), we can analytically find the power
\begin{equation}
P(\beta)=v_0+\frac{1+4\gamma v_0^2}{2\sqrt{\gamma}}tan^{-1}(2\sqrt{\gamma}v_0),\label{eq:nine}
\end{equation}
The derivative $dP/d\beta$ can be
gained from Eq.~(\ref{eq:nine}) easily and it transpires that the power monotonically increases with the propagation constant, indicating that the solitons are stable~\cite{28,29}.
To further demonstrate the stability of the exact soliton, we also do the direct numerical simulations of Eq.~(\ref{eq:six}) and find that the soliton is stable (Fig.~\ref{fig:one} (d)). Here, $v(x)$ is the exact soliton solution as an initial condition.

\section{Numerical solitons solutions}
\label{Numerical solitons solutions}
\begin{figure}[htb]
\centerline{\includegraphics[width=10cm]{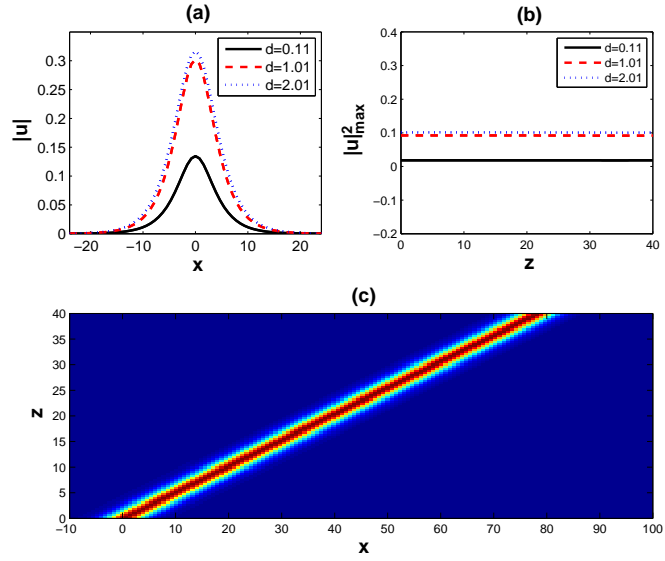}}
\caption{\label{fig:two}(Color online) (a) The profiles of the solitons. (b) Evolution of the peak value of the solitons power $|u|^2_{max}$.
(c) Simulated propagation of the beam at $d=0.11$ with $\rho=2\%$ noise for a numerical soliton solution initial condition. The other parameters $\kappa=1\times10^{-3}$ and $V=10$.}
\end{figure}

To discuss the influence of the degree of nonlocality and nonparaxiality on the width and stability of solitons, we use split step Fourier method (SSFM) and spectral renormalization method~\cite{27-1} to obtain the nonlocal nonparaxial soliton solutions by solving the equations (\ref{eq:three}), where a solution guess for field distribution is $u(x)=\sec$h$(x)$. Fig.~\ref{fig:two} (a) shows the soliton solutions with the different degree of nonlocality $d$. We find that the solitons power will increase with the degree of nonlocality $d$. However, as shown in Fig.~\ref{fig:two} (b), when $d=0.11$, $d=1.01$ and $d=2.01$, the peak values of the solitons intensity $|u|^2_{max}$ versus propagation distance are almost straight horizontal lines, that is to say, $d$ does not influence the stability of solitons. To illustrate the stability of the solitons, we do the direct numerical simulations of Eqs.~(\ref{eq:three}) with input condition $u|_{z=0}=v(1+\rho)\exp(-iVx)$, where $v(x)$ is the numerically nonlocal nonparaxial soliton solution as an initial condition. The result in Fig.~\ref{fig:two} (c) shows that the solitons can be stable when $d=0.11$, which is consistent with the exact soliton solution.

\begin{figure}[htb]
\centerline{\includegraphics[width=10cm]{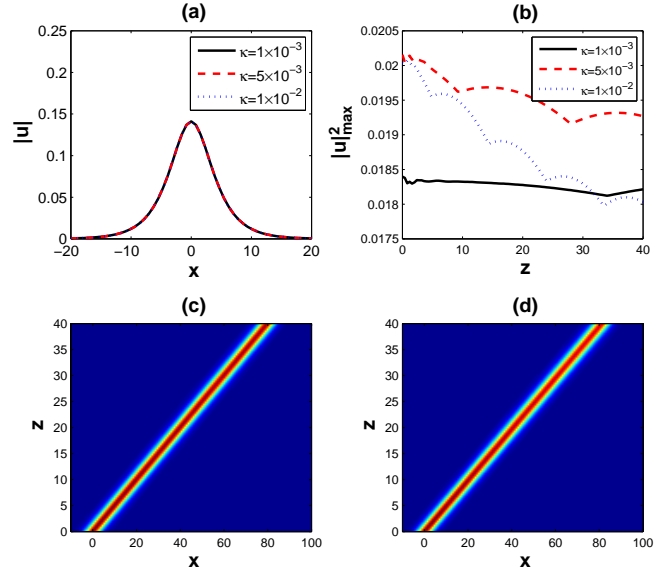}}
\caption{\label{fig:four}(Color online) (a) The profiles of the solitons. (b) Evolution of the peak values of the solitons power $|u|^2_{max}$.
(c) and (d) show the simulated propagations of the beams with $\rho=2\%$ noise at $\kappa=5\times10^{-3}$ and $\kappa=1\times10^{-2}$, respectively. The other parameters $V=10$ and $d=0.11$.}
\end{figure}

Next, we discuss the influence of non-paraxial parameter $\kappa$ on the solitons properties. From Fig.~\ref{fig:four} (a), we can also see that non-paraxial parameter $\kappa$ has no effect on the field distribution when $V$ and $d$ are constant. However, $\kappa$ can influence the stability of solitons. In Fig.~\ref{fig:four} (b), we easily find that the peak values of the solitons power $|u|^2_{max}$ decrease with the propagation distance when $\kappa=5\times10^{-3}$ and $\kappa=1\times10^{-2}$. The solitons will be no more stable when $\kappa=5\times10^{-3}$ and $\kappa=1\times10^{-2}$. Also, we get the same conclusions by comparing Fig.~\ref{fig:two} (c), Fig.~\ref{fig:four} (c) and (d), that is, the solitons will be no more stable when the non-paraxial parameter $\kappa$ is bigger. When $\kappa$ is bigger, the width of nonparaxial soliton beams $w_0$  should be smaller. The bigger $\kappa$ can introduce the first type of nonparaxiality, so the solitons are no stable.

\section{Conclusion}
\label{Conclusion}
In conclusion, we investigate the nonparaxial solitons in the nonlocal nonlinear medium. We find an exact analytical solution to the nonlocal nonlinear nonparaxial propagation equation using Taylor expansion, when nonlocality is high and weak. Moreover, we numerically find that the degree of nonlocality and nonparaxiality have different influence on the width and stability of solitons.

\section*{Acknowledgments}
This research is supported by the Natural Science Foundation of Guangdong Province, China (Grant S2012040007188).

 \clearpage
\listoffigures

\clearpage

\clearpage

\end{document}